\documentstyle[titlepage,12pt]{article}

\def\be{\begin{equation}}
\def\bea{\begin{eqnarray}}
\def\grad{\nabla}
\def\ee{\end{equation}}
\def\eea{\end{eqnarray}}

\newcount\sectionnumber
\def\sect
{\global\equationnumber=0
\global\advance\sectionnumber by 1
\the\sectionnumber . }
\newcount\equationnumber
\def   \num
{\eqno{\global\advance\equationnumber by 1
\left(\the\sectionnumber .\the\equationnumber \right)}}

\begin{document}
\title{Conservation Laws and 2D Black Holes in Dilaton Gravity}
\author{R.B. Mann \\
Department of Physics \\
University of Waterloo \\
Waterloo, Ontario \\
N2L 3G1}

\date{June 9, 1992\\
WATPHYS-TH92/03}

\maketitle

\begin{abstract}

A very general class of Lagrangians which couple scalar fields to
gravitation and matter in two spacetime dimensions is investigated. It is
shown that a vector field exists along whose flow lines the stress-energy
tensor is conserved, regardless of whether or not the equations of motion
are satisfied or if any Killing vectors exist. Conditions necessary for the
existence of Killing vectors are derived. A new set of 2D black hole
solutions is obtained for one particular member within this class of
Lagrangians. One such solution bears an interesting resemblance to
the 2D string-theoretic black hole, yet contains markedly different
thermodynamic properties.
\end{abstract}

\section{Introduction}

The study of gravitation in two spacetime dimensions has been a subject of
much study in recent years, motivated both by string theory and by a desire
to study quantum gravitational effects in a mathematically tractable
setting. Significant progress has been made in recent years as a
consequence of the realization that two-dimensional spacetimes with
non-trivial event horizons ({\it i.e.} black holes) exist
\cite{BHT,MST,MFound,EDW}.

The fundamental problem in constructing a relativistic theory of
gravitation in $(1+1)$ dimensions (2D) is the
triviality of the Einstein tensor: $G_{\mu\nu}[g]\equiv 0$ for all metrics
$g_{\mu\nu}$.  The most popular method of addressing this difficulty has
been to use the Polyakov action \cite{Polya}, a non-local action which
becomes the local Liouville action for a certain choice of coordinates.

However in the last few years it has been shown that other approaches exist
which are of significant interest in their own right. These approaches
typically couple a scalar field (the dilaton) to gravity in such a way that
the Ricci scalar is a non-vanishing function of the co-ordinates over some
region of spacetime, and so the spacetime can develop interesting features,
{\it i.e.} black-hole horizons and singularities. These approaches include
setting the Ricci scalar equal to a constant \cite{JackTeit}, setting the
Ricci scalar equal to the trace of the 2D stress-energy tensor
\cite{MST,MFound}, or setting the one-loop beta functions of the bosonic non-
linear sigma model to zero in a two-dimensional target space \cite{MSW}.
This latter approach yields a set of equations which are equivalent to
those derived from an effective action coupling the target space matric and
dilaton fields and is motivated from string theory \cite{EDW}. This action
has recently been extended to include more general dilaton couplings to
gravity and matter \cite{Frolov,Nappi}.

In the present paper a quite general class of actions which couple
a dilaton field to gravity and matter in two-dimensional spacetime is
analyzed. It will be shown that there always exists a flow line along which
the stress-energy tensor is conserved, regardless of the other equations of
motion in the theory or of the existence of a Killing vector. Conditions
necessary for the tangent vector to this flow line to be a Killing vector
are constructed.  These results are then
applied to the theories discussed in refs. \cite{MST,MFound,Frolov,Nappi}.
Finally, for one particular member in the class of Lagrangians considered
here, a new set of black hole solutions is derived and their thermodynamic
properties explored.

\section{Two-Dimensional Gravity and Dilaton Fields}

The action considered here is taken to be
\be
S[g,\Psi]
= \int d^2x\sqrt{-g}\left( H(\Psi)g^{\mu\nu}\grad_\mu\Psi
   \grad_\nu\Psi +D(\Psi)R + V(\Psi;\Phi_M)\right) \label{1}
\ee
where $R$ is the Ricci scalar and $D$ and $H$ are arbitrary functions of
the scalar field $\Psi$ (referred to as the dilaton in string theory). The
potential $V$ is the matter Lagrangian, and depends on both $\Psi$ and the
matter fields $\Phi_M$. This action is a generalization of one considered
recently by  Banks and O'Loughlin \cite{Banks}, in which no matter fields
$\Phi_M$ were present and $H=1/2$.

This model actually only depends upon the one function $V(\Psi;\Phi)$,
since reparametrizations of $\Psi$ accompanied by $\Psi$-dependent Weyl
rescalings of the metric allow one to relate models with different $H$'s
and $D$'s.  Explicitly, under the transformation
\be
g_{\mu\nu}=e^{\sigma(\tilde{\Psi})}\tilde{g}_{\mu\nu} \quad \Psi =
J(\tilde{\Psi}) \label{1a}
\ee
the action (\ref{1}) with $V=0$ is form
invariant provided $\tilde{D}(\tilde{\Psi})=D(J(\tilde{\Psi}))$ and
\be
(J^\prime)^2 \left(H(J)+ D^\prime(J) \sigma^\prime(J)\right) =
H(\tilde{\Psi}) \label{1b}
\ee
where
$\prime$ denotes the derivative with respect to the functional argument.
Only the critical points and overall sign of $H$ and $D$ contain
reparametrization invariant information \cite{Banks}.  In order for there
to be non-trivial evolution for the spacetime metric, it is necessary that
$D^\prime\neq 0$.

Of course a Weyl rescaling is information external to that derived from the
action (\ref{1}) via a variational principle. The matter potential $V$
breaks Weyl invariance, and so the field equations which follow from
(\ref{1}) determine the evolution of the spacetime metric and matter
fields. General coordinate invariance implies that locally the evolution of
the metric is determined by the evolution of its conformal factor.

The action (\ref{1}) is the most general action linear in the curvature and
quadratic in the derivatives of $\Psi$ and the matter fields. It
has been employed a number of times in constructing two-dimensional
theories of gravity.  For $H=0$ and $V=-D\Lambda$ this action reduces to
that considered in ref.\cite{JackTeit}.  For $H=1/2$, $D=\psi$ and  $V=-
8\pi G{\cal L}_M(\Phi)$, the action is that considered in
\cite{MST,MFound,semi}, in which the field equations set the Ricci
scalar equal to the trace of the stress energy.  For $H=\gamma e^{-2\phi}$
$D=e^{-2\phi}$ and  $V=-\frac{1}{4}e^{(\epsilon-2)\phi}{\cal L}_M(\Phi)$,
this yields the class of theories considered in \cite{Frolov,Nappi}.  For
$\gamma=4$ and $\epsilon=0$, the action reduces to that of the effective
target space action for non-critical string theory \cite{MSW,Frolov} the
matter action being that of the tachyon field.

\section{Conservation Laws and Killing Vectors}

Consider a potential $V$ such that
the stress-energy tensor associated with the action (\ref{1}) is
\be
T_{\mu\nu}\equiv\frac{1}{\sqrt{-g}}\frac{\delta S}{\delta g^{\mu\nu}}
= H\left(\grad_\mu\Psi\grad_\nu\Psi-\frac{1}{2}(\grad\Psi)^2\right)
-\grad_\mu\grad_\nu D + g_{\mu\nu}\grad^2D - \frac{1}{2}g_{\mu\nu}V
\quad , \label{2}
\ee
essentially restricting the potential to have no metric dependence. While
this does omit kinetic energy terms for scalars and spinors, it does permit
couplings to gauge field strengths since these are always dual to
scalars. The divergence of the stress energy is
\be
\grad^\mu T_{\mu\nu} = \frac{1}{2}\grad_\nu\Psi \left[
H^\prime(\grad\Psi)^2 -D^\prime R - V^\prime + 2 H \grad^2\Psi\right]
-\frac{1}{2} V^I\grad_\nu {\cal S}_I
\label{3}
\ee
where $\prime$ denotes the derivative with respect to $\Psi$ and
${\cal S}_I$ are scalar quantities formed from the matter fields
$\Phi_M$ and their derivatives, with $V^I\equiv \frac{\delta V}{\delta
S_I}$.   The right hand side of this equation will vanish when the equations
of motion of $\Psi$ and the matter fields  $\Phi_M$ are satisfied.

However even if these equations are not satisfied, the stress-energy tensor
obeys a conservation law.  Consider the divergence of the quantity
${\cal J}_\mu\equiv T_{\mu\nu}\epsilon^{\nu\alpha}\grad_\alpha F(\Psi)$:
\be
\grad^\mu{\cal J}_\mu
= \frac{1}{2}\epsilon^{\nu\alpha}\left[\grad_\nu\Psi
\grad_\alpha((\grad\Psi)^2)\right]\left(D^\prime F^{\prime\prime} -
F^\prime D^{\prime\prime} + H F^\prime\right)
-V^I\grad_\nu {\cal S}_I \epsilon^{\nu\alpha}\grad_\alpha F
\quad . \label{4}
\ee
The result (\ref{4}) holds regardless of whether $T_{\mu\nu}$ is conserved
or whether $\xi^\nu\equiv \epsilon^{\nu\alpha}\grad_\alpha F(\Psi)$ is a
Killing vector. It is clear that ${\cal J}_\mu$ is conserved provided
\be
\grad_\nu {\cal S}_I \epsilon^{\nu\alpha}\grad_\alpha F = 0
\label{5}
\ee
and
\be
F = F_0 \int^\Psi ds D^\prime e^{-\int^s dt \frac{H(t)}{D^\prime(t)}}
\quad , \label{6}
\ee
where $F_0$ is a constant.  This constant may be chosen so that
$\frac{dF}{dx}\to 1$ for large $|x|$.

The first of these conditions states that the gradients of the scalars
formed from matter fields and their derivatives ({\it e.g.} for gauge
fields the gradient of the dual of the gauge field strength) must be
orthogonal to surfaces of constant $F$.  The second condition
guarantees that $T_{\mu\nu}$ is always conserved along the flow lines
of $\xi^\nu$, even if $\grad^\mu T_{\mu\nu}\neq 0$.
Since in two dimensions a divergenceless current is always
dual to the gradient of a scalar, ${\cal J}_\mu = \epsilon_\mu^{\ \nu}
\grad_\nu{\cal M}$; from (\ref{2},\ref{6})
\be
{\cal M} = \frac{1}{2}\left[(\grad D)^2 e^{-\int^s dt
\frac{H(t)}{D^\prime(t)}}
- F_0 \int dD V e^{-\int^s dt \frac{H(t)}{D^\prime(t)}}\right]
\quad . \label{7}
\ee
The quantity ${\cal M}$ is a generalization of the mass-function $m(x)$
considered in ref. \cite{Frolov}. Note that ${\cal M}$ is constant whenever
the equation of motion for the metric ({\it i.e.} $T_{\mu\nu}=0$ in
(\ref{2})) is satisfied.

Consider next the conditions under which $\xi^\mu$ is a Killing vector.
For $H\neq D^{\prime\prime}$, (\ref{3}) and (\ref{6}) give
\be
\grad_{(\mu}\xi_{\nu)} = \frac{D^{\prime\prime}}{H}\grad_{(\mu}\xi_{\nu)}
+ \frac{F^{\prime\prime}}{H}T_{\alpha(\mu}\epsilon_{\nu)}^{\ \alpha}
\label{9}
\ee
and so,
\be
\grad_{(\mu}\xi_{\nu)} = - e^{-\int^\Psi dt \frac{H(t)}{D^\prime(t)}}
T_{\alpha(\mu}\epsilon_{\nu)}^{\ \alpha} \quad .
\label{10}
\ee
A similar computation when $H = D^{\prime\prime}$ (which is the case for
string theory \cite{MSW,Nappi}) yields
\be
\grad_{(\mu}\xi_{\nu)} = - \frac{F_0}{D^\prime}
T_{\alpha(\mu}\epsilon_{\nu)}^{\ \alpha} \quad .
\label{10a}
\ee
In either case it is clear that $\xi$ is a Killing vector only if
$T_{\mu\nu} = K(\Psi,\Phi_M)g_{\mu\nu}$.  This is the condition used in
ref. \cite{Banks} in deriving the general form
$\xi^\nu = \epsilon^{\nu\alpha}\grad_\alpha F$ (with $F$ given by (\ref{6}))
of the Killing vector associated with the action (\ref{1}) with $\Phi_M=0$.

To close this section, the above results will be applied to a few simple
examples.  For string theory $H=4e^{-2\phi} = 4D$ and $V=\lambda e^{-2\phi}$,
and
so (\ref{7}) becomes
\be
{\cal M} = -2 F_0 e^{-2\phi}\left( (\grad\phi)^2 - \frac{\lambda}{4}\right)
\quad . \label{11}
\ee
When the field equations are satisfied
\be
ds^2 = - (1- ae^{-Qx})dt^2 + \frac{dx^2}{1- ae^{-Qx}}  \label{12}
\ee
\be
\phi = -\frac{Q}{2}x \label{13}
\ee
where $a$ is a constant of integration and $\lambda=Q^2$. Normalizing $F_0$
as stated above yields
\be
{\cal M} = \frac{1}{2}aQ \label{14}
\ee
which is the expression for the ADM mass for the black hole obtained in
refs. \cite{EDW,Frolov,Nappi2}.  For the $R=T$ theory of
ref.\cite{MST,MFound}, $H=\frac{1}{2}$, $D=\psi$ and $V=0$ in the absence
of matter, and (\ref{7}) becomes
\be
{\cal M} = \frac{F_0}{2} (\grad\psi)^2 e^{-\psi/2} \quad  .\label{15}
\ee
Outside a symmetrically placed distribution of matter the exact
solution to the field equations is \cite{RGRG}
\be
ds^2 = -(2M|x|-1)dt^2 - \frac{dx^2}{(2M|x|-1)} \label{16}
\ee
and $\psi = -2\ln((2M|x|-1))$, and so ${\cal M}$ becomes
\be
{\cal M} = M \label{17}
\ee
where $F_0$ has been normalized as above.

\section{New Black Hole Solutions}

In this section the action
\begin{equation}
S=\int d^2x\,\sqrt{-g}\left(\frac{1}{2}
(\grad\psi)^2 + \psi R + 2b (\grad\phi)^2
- 8 \pi G\left[-f(\phi)\Lambda
+\frac{1}{4}h(\phi)F_{\mu\nu}F^{\mu\nu}\right] \right) \label{18}
\end{equation}
will be investigated, where $F_{\mu\nu}=\partial_\mu A_\nu-\partial_\nu
A_\mu$, and $f$ and $h$ are (at this point) unspecified functions of the
scalar field $\phi$.

This action is clearly a special case of (\ref{1}), and corresponds to the
R=T theory mentioned eariler \cite{MST,MFound}. Such a theory is of
interest because it yields a two dimensional theory which closely resembles
$(3+1)$ dimensional general relativity in that the evolution of the
gravitational field is driven by the stress-energy and no other Brans-Dicke
type fields \cite{RGRG}. Its classical and semi-classical properties and
solutions are also markedly similar
\cite{MST,semi,Arnold,Shardir,TomRobb}. Indeed, the field equations
which follow from (\ref{18}) may be obtainted from a reduction of the
Einstein equations from $D$ to 2 spacetime dimensions \cite{RGRG}.

These field equations are
\be
\grad^2\psi - R = 0  \label{20}
\ee
\begin{equation}
\frac{1}{2}\left(\grad_\mu\psi\grad_\nu\psi - \frac{1}{2} g_{\mu\nu}
(\grad\psi)^2\right) +
g_{\mu\nu}\grad^2\psi - \grad_\mu\grad_\nu\psi
= 8\pi G T_{\mu\nu} \label{21}
\end{equation}
\be
-4b\grad^2\phi + 8\pi G\left(\Lambda\frac{df}{d\phi}
-\frac{1}{4}\frac{dh}{d\phi}F_{\mu\nu}F^{\mu\nu}\right)
= 0 \label{22}
\ee
\be
\grad_\mu\left(h(\phi)F^{\mu\nu}\right)  = 0 \label{23}
\ee
where
\be
T_{\mu\nu} = \frac{1}{2}g_{\mu\nu}\Lambda f(\phi)
+\frac{1}{2}h(\phi)\left(F_{\mu\tau}F_{\nu}^{\ \tau}-
\frac{1}{4}F_{\rho\tau}F^{\rho\tau}\right)  \label{24}
\ee
is the stress-energy tensor due to the matter fields $\phi$ and $A_\mu$.

Taking the metric in the static case to be of the form
\begin{equation}
ds^2 = - \alpha(x) dt^2 + \alpha^{-1}(x) dx^2   \label{19}
\end{equation}
the gravity/matter system reduces to
\be
\alpha^{\prime\prime} = -8\pi G\Lambda f(\phi) + 4\pi G \frac{Q^2}{h(\phi)}
\label{25}
\ee
and
\be
\left(\alpha\phi^\prime\right)^\prime = \frac{2\pi G}{b}\Lambda f^\prime(\phi)
+\frac{\pi G}{b}\left(\frac{Q^2}{h(\phi)}\right)^\prime
\label{26}
\ee
where
\be
F_{\mu\nu} = \epsilon_{\mu\nu} \frac{Q}{h(\phi)}  \label{27}
\ee
is the solution to (\ref{23}). It is possible to use (\ref{25}) to
integrate (\ref{26}) when one of $Q$ or $\Lambda$ vanishes; the result is
\be
(\phi^\prime)^2 = \pm\frac{1}{4b}\frac{d}{d\alpha}\left[
\frac{(\alpha^\prime)^2 - X_0}{\alpha}\right] \label{28}
\ee
where the plus (minus) sign corresponds to $\Lambda=0$ ($Q=0$).
Equation (\ref{20}) always has the solution
\be
\psi = -\frac{\alpha^\prime+K}{\alpha} \label{29}
\ee
for the auxiliary field $\psi$.  Here $K$ and $X_0$ are constants of
integration.  Comparison with the 00 component of (\ref{21}) indicates
$2X_0 = K^2$.

Further progress requires specifying the functions $f$ and $h$. For
`clumped' matter, coordinates may be chosen so that the edges of the matter
are symmetric about the origin and located a finite proper distance away.
In this case one expects the metric to have a spatial dependence such that
$\alpha(x)$ is positive for large $|x|$. A black hole is a region of
spacetime for which $\alpha$ is negative; in the static case this will
occur for some range of $x$ \cite{DanRobb}.  This range may either be
finite, as in the case of the collapse of localized `clumped' matter
\cite{Arnold,symbh}, or infinite, as in the case of dilatonic black holes
\cite{MSW,EDW,Frolov}. It is also worthwhile to note that the criterion for
asymptotic flatness is slightly more general than in higher dimensions
\cite{RGRG}. One need only require that $\alpha(x)\to K|x|+C$ for large
$|x|$, (or perhaps just for large $x$) since in this case the metric
(\ref{19}) will become asymptotically like a Rindler spacetime; a Rindler
transformation may then be applied locally to  rewrite the metric in
Minkowskian form.

A particularly interesting class of static solutions to the system
(\ref{20})--(\ref{24}) results when $\phi$ is chosen to be a Liouville
field {\it i.e.} $f=e^{-2a\phi}$ and $h=f$.
For both $Q$ and $\Lambda$ non-vanishing there exists a solution
with $\beta=1$:
\be
\alpha(x) = A-\frac{8\pi G\Lambda}{C^2}e^{-2aE} e^{-Cx}
+\frac{4\pi GQ^2}{C^2}e^{2aE} e^{Cx}
\label{34}
\ee
\be
\phi= \frac{C}{2a}x+E  \label{33}
\ee
where $A$, $C$ and $E$ are constants of integration. This solution is not
asymptotically flat. However if either of $\Lambda$ or $Q$ is  non-vanishing,
a more physically interesting set of solutions results.

Consider first the case $Q=0$. Solving (\ref{25}) for $\phi$ and
substituting this into (\ref{28}) yields
\be
\beta \left(\alpha\frac{\alpha^{\prime\prime\prime}}{\alpha^{\prime\prime}}
\right)^\prime - \alpha^{\prime\prime} = 0
\ee
where $\beta\equiv -b/a^2$. There is a discretely infinite set of
solutions which are asymptotically Rindlerian at large $x$
whenever $\beta=\frac{p}{p+2}$
\be
\alpha(x) = C(x-x_0)-\frac{A_p}{(x-x_0)^{p}}  \label{30}
\ee
where $A_p$ and $C$ are constants of integration, and
\be
\phi = -\frac{1}{2a}\ln\left(\frac{p(p+1)A_p}{8\pi G\Lambda (x-
x_0)^{p+2}}\right) + \phi_0 \label{31}
\ee
For large $p$ ({\it i.e.} $\beta=1$)
the solution is
\be
\alpha(x) = A-\frac{8\pi G\Lambda}{C^2}e^{-2aE} e^{-Cx}
\label{32}
\ee
with $\phi$ the same as in (\ref{33}), and $A$, $C$, $E$ are
constants of integration. When $\Lambda=0$ and $Q\neq 0$, the solutions to
(\ref{20})--(\ref{24}) are given by (\ref{30})--(\ref{33}) with $a\to -a$
and $\Lambda$ replaced by $-Q^2/2$. For each of the solutions in
(\ref{34},\ref{32}), the constant $K^2=A^2C^2$ in (\ref{29}).

The temperature of the black hole solutions is straightforwardly obtained
by naive Wick-rotation arguments and is easily seen to be
$T=|\frac{d\alpha}{dx}/4\pi|_{x_H}$ where $x_H$ is the location of the
horizon ($\alpha(x_H)=0$).  For (\ref{30}) this is
\be
T = \frac{C(p+1)}{4\pi}                        \label{35}
\ee
whereas for (\ref{32}) this is
\be
T = \frac{C}{4\pi}                        \label{35a}
\ee
whenever $C$ and $A_p$ are of appropriate relative sign so that an event
horizon exists.

\section{Discussion}

The solutions (\ref{30},\ref{32}) represent an interesting new class of 2D
metrics. For large $x$ they are asymptotic to either  Rindler space (as in
(\ref{30})) or flat space (as in (\ref{32}))  and for appropriate choices
of the signs of $C$, $A_p$, they have black hole event horizons.  If $x$ is
replaced by $|x|$ in (\ref{30},\ref{32}) then they remain solutions of the
system (\ref{20})--(\ref{24}) provided an appropriate point-source stress-
energy \cite{MST} is inserted at the origin. Such metrics would represent
the endpoint of gravitational collapse of localized matter. Due to the
presence of the $\phi$ field, a full treatment of such a problem would
involve consideration of matching conditions at the boundary
\cite{symbh,rosscoll}. For (\ref{32}) the curvature tensor would have a
delta-function singularity at the origin, and a power-law singularity at
$x=x_0$.

The solution (\ref{32}) is of particular interest. Formally, it is
identical to the string-theoretic solution (\ref{12}).
However its interpretation is quite different. The
coefficient $C$ inside the exponential in (\ref{32}) is a constant of
integration, in contrast to the $Q$ parameter in (\ref{12})
which is a coupling parameter of the theory. The ADM mass computed from
(\ref{12}) is proportional to $aQ$, leading one to interpret $a$ as a mass
parameter (although an alternate interpretation has been proposed
\cite{symbh}), whose dimensionality is given by the fundamental coupling $Q$.

However for the theory described by (\ref{18}), the constant of integration
in (\ref{32}) is $C$. This suggests that it ought to be interpreted as a
mass. In computing the ADM mass associated with (\ref{32}), the formula
(\ref{7}) cannot be used, since the $V$ term in (\ref{18}) explicitly
depends on the metric.  However one can proceed directly, using the fact
that $\xi^\mu=(1,0)$ is a Killing vector for the spacetime (\ref{32});
defining the mass-function $M$ analogously to (\ref{7}) so that
$\grad_\mu M = -\epsilon_{\mu\nu}T^\nu_\rho \xi^\rho$
one obtains from (\ref{29},\ref{33})
\be
M= \alpha^\prime -\frac{b}{a}\alpha\phi^\prime -\int\left[\frac{1}{4}
\left(\frac{(\alpha^\prime)^2-K^2}{\alpha}\right)
-b\alpha(\phi^\prime)^2\right]  \label{36}
\ee
and using (\ref{32}) this becomes
\be
M =\frac{AC}{2} \label{37}
\ee
which upon normalizing (\ref{32}) so that $A=1$, yields $C=2M$.

This intepretation of the mass parameter is in striking contrast to that
considered in string theoretic contexts \cite{EDW,MSW,Frolov} and leads
to significantly different thermodynamic properties for the black hole
described by (\ref{37}). In contrast to the metric (\ref{12}) in which the
temperature is independent of the black hole mass, and the entropy is
proportional to it, in this case the temperature (\ref{35a})
varies linearly with the mass and the entropy logarithmically.  Such an
interpretation is in keeping with previous results in the R=T theory
\cite{MST,TomRobb}. Consequences of a black hole entropy S for which
$S\sim \ln(M)$ have been investigated previously \cite{TomRobb}. The
implications of this for black hole evaporation remain to be explored.

\section*{Acknowledgements}

This work was supported by the Natural Sciences and Engineering Research
Council of Canada.

\end{document}